\documentstyle[epsf]{mn}

\def\plotone#1{\centering \leavevmode
\epsfxsize=\columnwidth \epsfbox{#1}}

\def\plotlong#1{\centering \leavevmode
\epsfxsize=1.4\columnwidth \epsfbox{#1}}

\title[Timing Measurements and Their Implications for Four Binary 
Millisecond Pulsars]{Timing Measurements and Their Implications for Four Binary Millisecond Pulsars}

\author[J.~F.~Bell]{J.~F.~Bell$^{1,5}$, M.~Bailes$^2$, R.~N.~Manchester$^3$, 
A.~G.~Lyne$^1$, F.~Camilo$^1$ and 
\newauthor J.~S.~Sandhu$^4$ \\
$^1$The University of Manchester, NRAL, Jodrell
       Bank, Macclesfield, Cheshire SK11~9DL, UK. \\
$^2$Physics Department, University of Melbourne,
       Parkville, Victoria 3052, Australia. \\
$^3$Australia Telescope National Facility, CSIRO,
       PO Box 76, Epping, NSW 2121, Australia. \\
$^4$Department of Astronomy, 105-24, California
Institute of Technology, Pasadena, CA 91125, USA \\
$^5$E-mail: jb@jb.man.ac.uk}

\begin{document}
\maketitle

\begin{abstract}

We present timing observations of four millisecond pulsars, using data
obtained over three years at the ATNF Parkes and NRAL Jodrell Bank radio
telescopes.  Astrometric, spin, and binary parameters are updated, and
substantially improved for three pulsars, PSRs~J0613$-$0200, J1045$-$4509
and J1643$-$1224.  We have measured the time variation of the projected
semi-major axis of the PSR~J0437$-$4715 orbit due to its proper motion, and
use it to constrain the inclination of the orbit and the mass of the
companion.  Some evidence is found for changes in the dispersion measures of
PSRs~J1045$-$4509 and J1643$-$1224. Limits are placed on the existence of
planetary mass companions, ruling out companions with masses and orbits
similar to the terrestrial planets of the solar system for eight
pulsars. \end{abstract}


\section{Introduction}

Millisecond pulsars (MSPs) can be used for a variety of astrophysical
applications and tests of physics. They are sensitive probes of the
interstellar medium \cite{bhvf93}.  They provide information about the
evolution of binary systems and white dwarfs \cite{bv91}, and have
unsurpassed sensitivity when used to search for planetary mass bodies
\cite{wol94}.  As excellent clocks, they challenge the stability of
terrestrial time standards in the best cases \cite{tay91}, place limits upon
the energy density of cosmological gravitational waves \cite{rt83,ktr94},
and are used in testing theories of relativistic gravity \cite{twdw92}.

Millisecond pulsars are generally weak radio sources, and their short
periods make them difficult to discover.  Over the last five years several
surveys have increased from four to 35, the number of millisecond pulsars
known in the Galactic disk \cite{cam96}, more than half of these new
discoveries resulting from the Parkes southern sky survey (Manchester et
al. 1996; A.~G. Lyne et al. in prep.)
\nocite{mld+96}.

In this paper we report on the timing results obtained from observations of
four millisecond pulsars with data spans of over 1000\,d.
PSRs~J0437$-$4715, J0613$-$0200, J1045$-$4509 and J1643$-$1224 were
discovered during the Parkes survey for millisecond pulsars and, with the
exception of PSR~J0437$-$4715 \cite{bbm+95}, have had only initial timing
parameters reported, based on a data span of approximately one year
\cite{bhl+94,lnl+95}.  The observations and timing analysis are
described in Section \ref{s:obs} while the timing residuals, pulsar
parameters, and pulse profiles are presented in Section
\ref{s:prp}. Observations of the time variation of the projected semi-major
axis of the orbit of PSR~J0437$-$4715 is reported in Section \ref{s:xdot}.
In Section \ref{s:pbe} we summarise all measurements of orbital
eccentricities for binary pulsars while, in Section \ref{s:plc} we use
Lomb-Scargle spectra to assess the timing residuals for evidence of
planetary mass companions.

\section{Observations and Timing Analysis}
\label{s:obs}

From the time of their discovery until 1996 March, at intervals averaging
about three weeks, pulse arrival times for the four MSPs were obtained using
the ATNF Parkes radio telescope with cryogenic receivers at centre
frequencies of 436, 660, 1520, 1940 and 2320\,MHz. To date, only relatively
small datasets have been obtained at 660, 1940 and 2320\,MHz.  Orthogonal
linear polarizations were observed using a $2\times256\times0.125$-MHz
filter bank at 436 and 660\,MHz while both hands of circular polarization
were observed with a $2\times64\times5.0$\,MHz filter bank at 1520, 1940 and
2320\,MHz.  After detection, the signals from the two polarizations were
added, one-bit digitised at appropriate sampling intervals (0.3\,ms at 436
and 660\,MHz, 0.1\,ms at 1520, 1940 and 2320\,MHz, 0.08\,ms for
PSR~J0437$-$4715) and written to magnetic tape.  Off-line, the data were
folded at the topocentric pulsar period to produce mean pulse profiles for
each frequency channel, typically with integration times of 90--180\,s.
These profiles were then transformed to the Fourier domain, phase-shifted
appropriately to compensate for dispersive delays across each observing
band, transformed back to the time domain and summed to form a final mean
profile at each frequency.

Pulse arrival times for two of the pulsars, J0613$-$0200 and J1643$-$1224,
have also been obtained using the 76-m Lovell telescope at Jodrell Bank
with cryogenic receivers at 408, 606, and 1404\,MHz.  Both hands of
circular polarization were observed using a $2\times64\times0.125$-MHz
filter bank at 408 and 606\,MHz and a $2\times32\times1.0$-MHz filter
bank at 1404\,MHz.  After detection, the signals from the two polarizations
were added, filtered, digitised at appropriate sampling intervals,
dedispersed in hardware before being folded on-line, and written to disk.

For both data sets, a standard pulse template was fitted to the observed
profiles at each frequency to determine the pulse times-of-arrival (TOAs).
The TOAs, weighted by their individual uncertainties determined in the
fitting process, were analysed with the {\tt TEMPO} software package
\cite{tw89}, using the DE200 ephemeris of the Jet Propulsion Laboratory
\cite{sta82} and the Blandford \& Teukolsky (1976)\nocite{bt76} timing model
for binary pulsars.  Using the measured TOAs and an initial set of
parameters describing the pulsar system, {\tt TEMPO} minimizes the sum of
weighted squared {\it timing residuals}, the difference between observed and
computed TOAs, yielding a set of improved pulsar parameters and post-fit
timing residuals. The dispersion measures were determined by measuring the
delays between the arrival times obtained at Parkes at two frequencies. We
also fitted offsets between all other sets of data for each pulsar, to
account for the different pulse shapes for each observing system and at
different frequencies.

\section{Timing Residuals, Pulsar Parameters, and Pulse Profiles}
\label{s:prp}

\begin{figure}
\plotlong{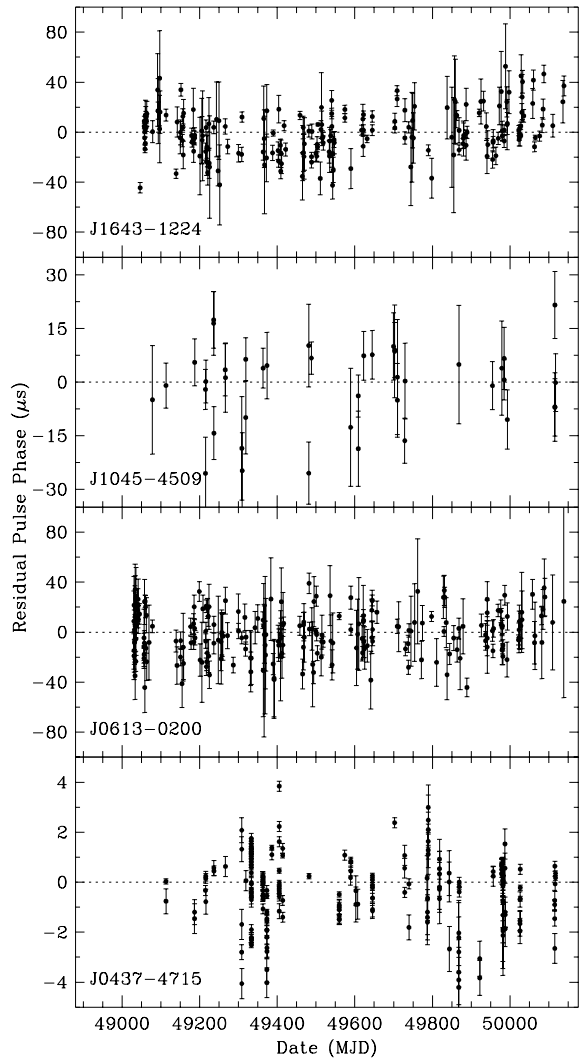}
\caption{Post-fit timing residuals for four MSPs plotted against MJD.}
\label{f:mjd}
\end{figure}

The parameters obtained from fits to the data for each pulsar are shown in
Table~\ref{t:par}.  The uncertainties quoted for these parameters are twice
the formal values reported by {\tt TEMPO}.  All parameters shown in
Table~\ref{t:par} were included in a global fit, except for $\ddot{P}$,
which was fitted for separately. The dispersion measure variation of
PSR~J1045$-$4509 was also obtained from a separate fit.  For
PSR~J0437$-$4715, only 1520 to 2320-MHz data were included in the global fit
due to the much larger TOA uncertainties at lower frequencies arising from
the greater dispersion smearing and scintillation.  In the case of
PSR~J1045$-$4509, only the 1520-MHz data were included in the global fit due
to the nonmonotonic dispersion measure variation with time.  For the other
two pulsars all available data were used.  Post-fit timing residuals are
shown for the four MSPs as a function of time in Fig.~\ref{f:mjd}. All radio
frequencies are plotted with the same symbol since there are no
frequency-dependent trends in the residuals.

For PSRs~J0613$-$0200, J1045$-$4509 and J1643$-$1224 the newly obtained
celestial coordinates and binary parameters, in particular the
eccentricities, are considerably improved over previously published values
and are discussed in more detail in Section \ref{s:pbe}.  The uncertainties
in the proper motion for PSR~J0437$-$4715 are improved by a factor of 10
while useful limits are obtained on the proper motions of the other three
pulsars.  The velocities shown in Table~\ref{t:par} were obtained using
distances calculated with the Taylor \& Cordes (1993)
free-electron-distribution model\nocite{tc93}.  For PSR~J0437$-$4715, the
upper limit on the parallax of 6 \,mas constrains the distance to this
pulsar to be greater than 166\,pc ($1 \sigma$), 90\,pc ($2 \sigma$),
compared to the dispersion measure distance of $140$\,pc with a 30 per cent
uncertainty. Assuming that the intrinsic period derivative is zero, so that
the measured period derivative is entirely due to the time-varying Doppler
effect caused by the proper motion \cite{shk70,ctk94}, gives an upper limit
on the distance of 208\,pc. The upper limits on $\ddot{P}$ for the four
pulsars provide no evidence for timing noise detectable with the current
level of timing precision and data spans.

In some cases, deviations of the post-fit timing residuals from zero are
somewhat larger than would be expected from the independently estimated TOA
uncertainties.  We believe that the TOA uncertainties are slightly
underestimated, and we account for this in the final fitting procedure by
multiplying the nominal TOA uncertainties by a small factor (typically
approximately two) so as to produce a reduced $\chi^2$ for the fit of
unity. The original uncertainties are shown in the figures.

In the case of PSR~J0437$-$4715 these deviations are very large
(Fig.~\ref{f:mjd}), and we believe they have a systematic origin.  From one
observation of this pulsar to another the ratio of the main peak intensity
to that of its outlying components appears to vary by up to
20 per cent, while the ratio of the component immediately following the main
peak to the underlying emission varies by factors as large as two.  We do
not know the exact cause of these variations, but suspect that the observed
changing pulse shapes may be caused by a relative gain variation in the two
polarization channels and the high degree of polarization of the emission
observed across the profile \cite{mj95}.  To obtain the best set of profiles
for timing, the profiles for different days were aligned and summed using
the best available ephemeris.  Each profile was then normalised and
subtracted from the sum.  The resulting difference profiles were used to
select a sample of profiles most similar in shape to the mean profile, and
were subsequently used to obtain the timing solution presented.  In making
this selection, some time-span has been sacrificed but the timing accuracy
has been significantly improved.  Nevertheless, the timing precision
obtainable for PSR~J0437$-$4715 with these data is not yet sufficient to
measure parameters such as parallax, or the apparent orbital period
derivative caused by the large proper motion of the pulsar \cite{bb96}.

\begin{figure}
\plotone{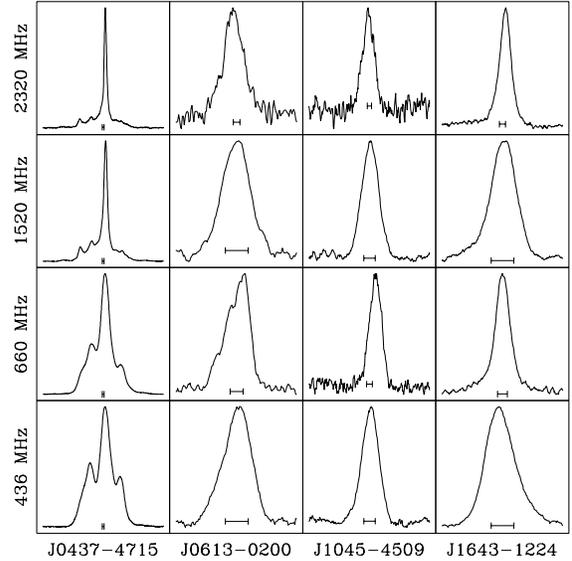}
\caption{Integrated pulse profile for each pulsar at four frequencies.  In
each case the full period is shown and the instrumental resolution is
indicated by a horizontal bar.}
\label{f:profs}
\end{figure}

Integrated pulse profiles for each of the pulsars are shown in
Fig.~\ref{f:profs} at four frequencies.  In all cases the intrinsic pulse
widths are narrower than the instrumental resolution. At 660\,MHz the
resolution for all four pulsars is determined by the sampling interval, as
is the resolution of the PSR~J0437$-$4715 profiles at 436 and 1520\,MHz.
The resolution of the remaining profiles is dominated by dispersion smearing
across individual channels.  Hence, for these pulsars, the profiles observed
at 1520 and 436\,MHz are broader than the corresponding profiles at
2320 and 660\,MHz. The asymmetry and tail of the PSR~J1643$-$1224
profile at 436\,MHz suggests some evidence for scattering along the line of
sight to this pulsar. The scattering time-scale $\tau_{s}$ \cite{tml93} as
measured from the pulse profile in Fig. \ref{f:profs}, is approximately
240\,$\mu$s at 436\,MHz. After scaling to 1\,GHz by $\nu^{-4.4}$
\cite{cwb85}, $\tau_{s,{\rm 1 GHz}} = 6 \mu$s and is comparable to the
scattering time-scales of other pulsars with similar dispersion measures
\cite{tml93}. It is however, somewhat higher than the $0.7 \mu$s predicted
by the Taylor \& Cordes (1993) model.

\begin{figure}
\plotone{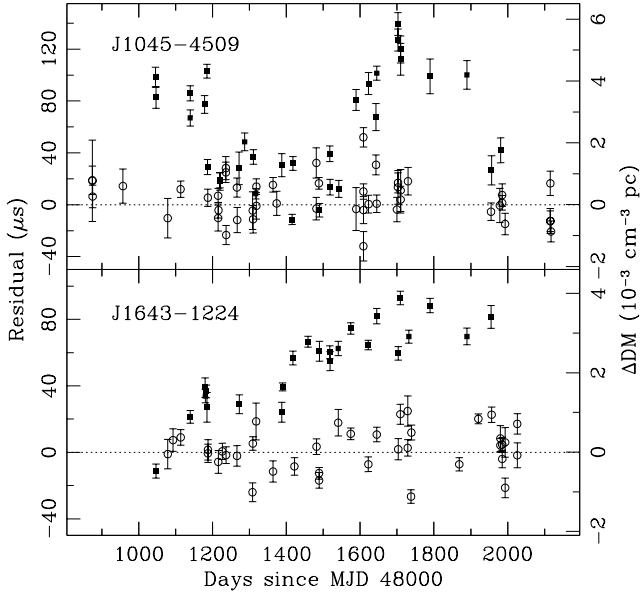}
\caption{Dispersion measure variations in two pulsars.  Timing
residuals from Parkes data at 436\,MHz (filled squares) and 1520\,MHz
(circles), relative to a timing model fitted to the latter.  Note the
systematic variations in the difference between the two frequencies.}
\label{f:ddm}
\end{figure}

We have observed dispersion measure variations for PSRs~J1045$-$4509 and
J1643$-$1224 and obtained an upper limit on the variations for
PSR~J0613$-$0200 as shown in Table~\ref{t:par}.  In the case of
PSR~J1045$-$4509, $\dot{\rm DM}$ was fitted over the first half of the data
span only, where the variation is monotonic.  To illustrate the variations
for PSRs~J1643$-$1224 and J1045$-$4509, the best fit to the Parkes 1520-MHz
data was made. For PSR~J1045$-$4509 the DM was fitted around MJD 49500 and
for PSR~J1643$-$1224 the DM was fitted around MJD 49100. The residuals for
the 436-MHz Parkes data where then calculated and plotted with the 1520-MHz
data in Fig.~\ref{f:ddm}.

Backer~et~al. (1993) \nocite{bhvf93} summarized observations of dispersion
measure variations for 13 pulsars and found that the rate of change of
dispersion measure, $\Delta DM \propto \sqrt{DM}$. This is somewhat flatter
than the linear dependence observed for diffractive scattering
\cite{cpl86}.  Dispersion measure variations sample electron-density
fluctuations on length scales of $10^{13}$--$10^{15}$\,cm, which are
$10^{4}$ times larger than the scale of the fluctuations which lead to
diffractive scattering.  This difference in variation with dispersion
measure suggests that these two effects may occur at different locations
along the line of sight, or have different driving mechanisms. Our results
are in general agreement with $\Delta DM \propto
\sqrt{DM}$ and therefore the conclusions drawn by Backer et al., although the
scatter of the data about this relation is large.

\section{Secular Variation of Orbital Inclination}
\label{s:xdot}

It has been recently noted that the proper motion of a pulsar can lead to a
secular variation of the projected semi-major axis of the pulsar's orbit,
$a_p \sin i$, as the orbital inclination $i$ changes
\cite{ajrt96,kop96,sbm+96}.  The contribution to the time derivative of the
projected semi-major axis is
\begin{equation}
\dot x = 1.54\times10^{-16} x\,\mbox{cot}\,i (-\mu_{\alpha} \sin \Omega +
\mu_{\delta} \cos \Omega),
\label{e:xdot}
\end{equation}
where $x=a_p \sin i/c$ is the projected orbital light travel time in
seconds, $\Omega$ is the longitude of the ascending node of the orbit, and
$\mu_{\alpha}$ and $\mu_{\delta}$ are the proper motions in right ascension
and declination, in mas\,yr$^{-1}$ \cite{kop96}.  Both the proper motions
can be measured, while $\Omega$ cannot, so that a measurement of $\dot x$
allows a constraint to be placed on $i$ and hence the mass of the companion
and on many aspects of the evolution of these systems.

The types of system for which this method works well are those that have
wide, small inclination (i.e. face-on) orbits, that are relatively nearby.
It is therefore complementary to the measurement of the ``Shapiro delay'',
which produces measurable effects for high-inclination (edge-on) systems in
tight orbits.  For millisecond pulsars with white-dwarf companions, the
orbit needs to be very close to edge on for the Shapiro delay to be
measurable. For intermediate inclination angles, where some evidence of
Shapiro delay is visible but where the covariance between the fitted
companion mass and inclination angle is great (e.g. in the PSR~J1713+0747
system, with $i \sim 70^\circ$, Camilo, Foster, \& Wolszczan
1994\nocite{cfw94}), measurement of $\dot x$ may be possible and should
provide an additional constraint on the system.

From the observations of PSR~J0437$-$4715, the measured value of $\dot
x=(7.4\pm1.4)\times 10^{-14}$ implies that $\dot x >6 \times 10^{-14}$.  Due
to the non-Gaussian distribution of the timing residuals, we used bootstrap
resampling \cite{ptvf92} to derive the uncertainties on all parameters in
Table \ref{t:par}, including $\dot{x}$. For most parameters, this yields
similar uncertainties to those calculated in the standard way from a
least-squares fit. However, for PSR~J0437$-$4715, the uncertainties on
$\dot{x}$ and $x$ are eight times larger using bootstrap resampling. We
therefore take the more conservative bootstrap uncertainties as the 68 per cent
confidence level. The maximum contribution from the proper motion occurs
when the proper motion is aligned with the semi-major axis of the orbit
projected onto the plane of the sky. This yields a limit of $\mbox{cot}\,i >
6.5\times10^{15}\dot x/(x
\mu) = 0.83$, giving an upper limit of $i < 50^{\circ}$. Consequently, the
companion mass is constrained to be greater than 0.19\,M$_\odot$, for an
assumed pulsar mass of 1.4\,M$_\odot$, considerably more than the {\it a
priori} minimum possible value of 0.14\,M$_\odot$.

One may easily show that the measured $\dot x$ cannot be due to a
genuine shrinking of the orbit.  Since $\dot x = \dot{a}_p/c \sin i +
a_p/c \cos i~{\rm d} i /{\rm d} t$, one must ensure that $\dot{a}_p
\sin i \ll a_p \cos i {\rm d} i /{\rm d} t$.  This can be checked by
obtaining an observational limit on $\dot{P}_{b}$ and differentiating
Kepler's third law with respect to time, which gives $\dot{a}_p$ as a
function of $\dot{P}_{b}$.  Inserting the observed limit for
PSR~J0437$-$4715 of $\dot{P}_{b} < 2.2 \times 10^{-11}$ shows that
$\dot x < 3.1 \times 10^{-15}$ which is indeed much smaller than the
measured value of $\dot x$, and hence shows that $\dot x \simeq a_p/c
\cos i {\rm d} i /{\rm d} t$.  The negligible value of $\dot{a}_p$ is
exactly as expected since orbital evolution due to gravitational
radiation or tidal effects leads to values of $\dot{a}_p$ many orders of
magnitude smaller than observed for $\dot x$.

\section{Orbital Eccentricities}
\label{s:pbe}

The timing observations presented here, vastly improve the accuracy of
the eccentricity measurements, particularly for PSRs J0613$-$0200 and
J1643$-$1224. In this section, we collate all such measurements of, and upper
limits on, eccentricities of binary pulsar orbits and compare them with
theoretical predictions.

\begin{figure}
\plotone{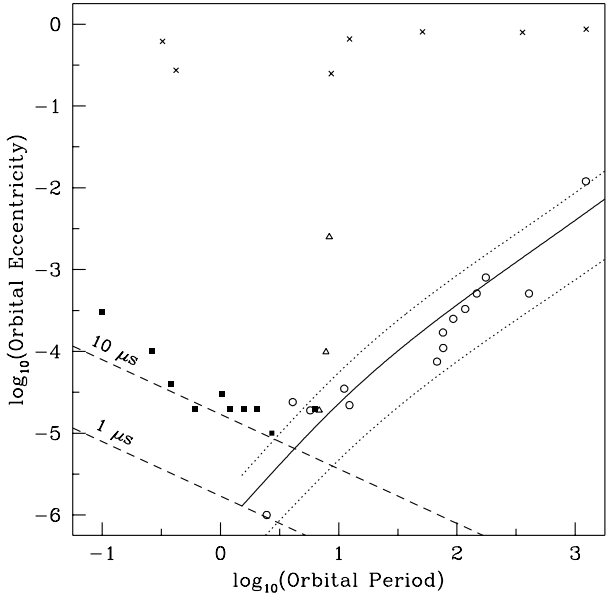}
\caption{Orbital eccentricities from Table~2 of Camilo
(1996)\protect\nocite{cam96} and this paper.  Solid line --- the predicted
median eccentricity for pulsars recycled by stable mass transfer from a
Roche-lobe-filling, low-mass red giant.  Dotted lines --- 95 per cent of the
measured eccentricities for recycled pulsars are expected to lie within
these \protect\cite{phi92b,pk94}.  Circles --- low-mass binary
pulsars. Squares -- upper limits on measured eccentricities.  Crosses --
high-mass binary pulsars.  Triangles - intermediate-mass binary
pulsars. Dashed lines -- approximate limits obtainable on eccentricities for
timing residuals of $10\,\mu$s and $1\,\mu$s.}
\label{f:pbe}
\end{figure}

The convective fluctuation-dissipation theory of Phinney (1992)
\nocite{phi92b} predicts a strong correlation between the orbital
eccentricities and orbital periods of binary pulsars that have been recycled
by stable mass-transfer from a Roche-lobe-filling, low-mass red giant.  The
measured eccentricities together with upper limits are shown for all
Galactic binary pulsars in Fig.~\ref{f:pbe}.  All measurements (circles)
and upper limits (squares) for objects with appropriate evolutionary
histories are in excellent agreement with the model.  As shown by the dashed
lines, the trend of the upper limits for those objects with short orbital
periods is merely a reflection of the timing precision attainable.

Neutron stars in highly eccentric orbits have provided some remarkable tests
of general relativity and other theories of gravity \cite{twdw92}.  There
are also interesting and useful tests than can be done using neutron stars
in very circular orbits: tests of the strong equivalence principle
\cite{ds91}, of local Lorentz invariance \cite{de92a} and of conservation
laws \cite{bd96}. These latter two tests depend linearly on the orbital
eccentricity of a sample of binary millisecond pulsars.

The orbit with the lowest eccentricity in Fig.~\ref{f:pbe}, and in
fact the known universe, is PSR~J2317+1439 \cite{cnt96}.  This has been
used to obtain a very tight limit on the local Lorentz invariance of
gravity \cite{bcd96}.  However, due to an unfavourable orientation of
its position and proper motion with respect to the dipolar structure of
the cosmic microwave background, the limit is not as tight as it might
otherwise be.  The predicted eccentricities for some millisecond pulsars
(e.g. PSR~J0613$-$0200) are somewhat lower than that of
PSR~J2317+1439, suggesting that these systems may ultimately provide
more stringent tests of local Lorentz invariance, and may also allow
the tests of the strong equivalence principle and conservations laws to
be improved.  However, as Fig.~\ref{f:pbe} shows, this will
require timing precisions of about 1\,$\mu$s, which will be very
difficult to achieve.  Nevertheless, any improvement on timing precision
for the pulsars with only upper limits on eccentricity will provide further
tests of the fluctuation-dissipation theory.

\section{Limits on Planetary Mass Companions}
\label{s:plc}

The discovery of planetary mass companions to a millisecond pulsar
\cite{wf92} was unexpected.  Their existence has been confirmed by the
detection of three-body perturbations in the timing residuals
\cite{wol94}.  Naturally, such a discovery stimulates interest in finding
other MSPs with planetary mass companions.  In this Section, we place limits
on the existence of any planetary mass companions to the pulsar discussed in
this paper using Lomb-Scargle spectra.  While these pulsars have
stellar-mass companions and may be less likely to have planetary companions,
we apply the same technique to four single pulsars in another paper
(M. Bailes et al. in prep.).

From the residuals shown in Fig.~\ref{f:mjd} there appears to be no
evidence for modulations that could be attributed to planetary mass
companions. The mass function for a pulsar binary is given by
\begin{equation}
f = {(m_{c} \sin i)^{3} \over (m_{p} + m_{c} )^{2} } = {4 \pi^{2} ( a_{p}
\sin i )^{3} \over G P_{b}^{2}}
\label{e:mfunc2}
\end{equation}
where $i$ is the inclination angle of the orbit, $a_{p}$ is the semi-major
axis of the pulsar's orbit, $P_{b}$ is the orbital period, $G$ is the
gravitational constant and $m_{p}$ and $m_{c}$ are the pulsar and companions
masses.  Assuming circular orbits and $m_{c} \ll m_{p}$, rearranging equation
\ref{e:mfunc2} suggests upper limits on the mass of undetected companions
of
\begin{eqnarray}
m_{c} \sin i & = & 
a_{p} \sin i \left( 4 \pi^{2} m_{p}^{2} \over G P_{b}^{2} \right)^{1/3} \\
& \leq & 
0.0143 M_{\oplus} \Delta T_{a} \left( P_{b} \over 1\,\mbox{day} \right)^{-2/3} 
\label{e:pmass}
\end{eqnarray}
where $\Delta T_{a}$ is the timing signal (peak-peak delay) due to the light
travel time across the pulsar's orbit and $M_{\oplus}$ is the mass of the
Earth.  If no modulation is seen in the residuals, a limit on $\Delta T_{a}$
of an unseen companion can be obtained from the variance of the residuals,
allowing a limit to be placed on $m_{c} \sin i$.  For long orbital periods
the sensitivity drops as $P_{b}^{3}$ due to the strong covariance between
the orbital period of a planet and the period derivative of the pulsar
\cite{tp92}.

The above limits should be considered as qualitative.  Quantitative limits
on the presence of modulations can be obtained through a number of
approaches: fast Fourier transforms (FFTs), fold and average, fitting of
orbits \cite{tp92}, and Lomb-Scargle spectra \cite{lom76,sca82,ptvf92}.
Each of these methods has various advantages and disadvantages which are now
briefly discussed.  Fold and averaging, and fitting orbits to residuals are
computationally expensive, provide poor detectability and sensitivity and do
not offer a statistical assessment of the significance of any signals
detected.  FFTs overcome all of these difficulties, but suffer from aliasing
and poor sensitivity to long orbits resulting from the need to fill in gaps
in the unevenly sampled data.

The Lomb-Scargle periodogram overcomes these difficulties and is similar in
speed to the FFT when one accounts for the additional data points required
to provide the even sampling for the FFT.  It also allows one to search for
frequencies up to several times the mean Nyquist frequency.  However, both
the Lomb-Scargle periodogram and the FFT fail to take account of the
covariance with other effects such as the Earth's orbit, which causes a loss
of sensitivity for periods around one year \cite{tp92}. A distinct advantage
of the spectral methods is that they do allow a formal assessment of the
statistical significance of spectral features, taking account of the number
of frequencies that were searched.  An ideal analysis therefore might
consist of a Lomb-Scargle periodogram, combined with a small selection of
fitted orbits to account for the above covariance problems.

\begin{figure}
\plotlong{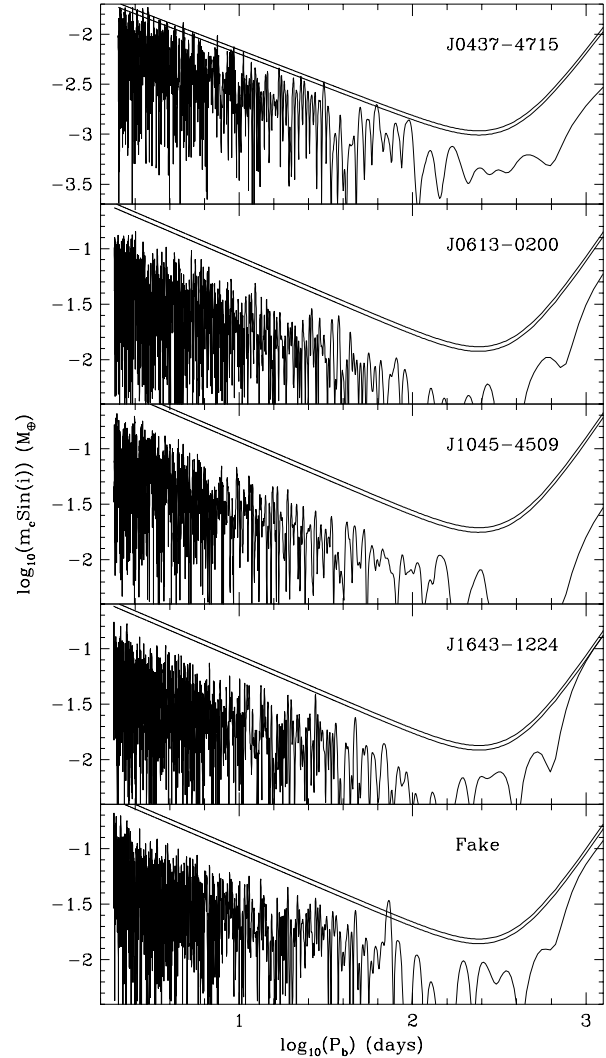}
\caption{The top four panels show Lomb-Scargle spectra of timing residuals
for pulsars in this paper.  In each panel the two smooth curves indicate the
0.1 and 0.999 probabilities that a signal is due to random noise. Thus
features above the line are likely to be significant signals and those below
the line, noise. The lower panel shows the result of modifying the
eccentricity of PSR~J1643$-$1224 to introduce a fake sinusoidal signal at
half the orbital period.  A significant detection is seen at precisely this
period.}
\label{f:plc}
\end{figure}

\begin{figure}
\plotlong{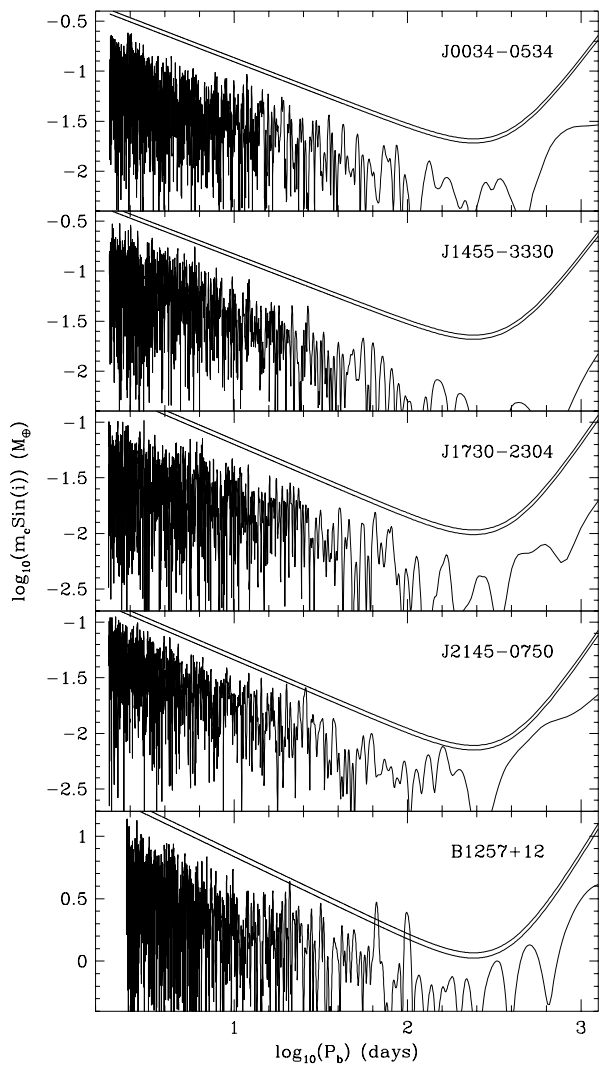}
\caption{Lomb-Scargle spectra of timing residuals for the four pulsars
discussed by Camilo~et~al. (1996)\protect\nocite{cmb+97}, as for
Fig.~\protect\ref{f:plc}.  In this case the lower panel shows a spectrum of
the Jodrell Bank timing residuals for PSR~B1257+12 which is known to have
planets.}
\label{f:plc2}
\end{figure}

In this paper, examples of Lomb-Scargle spectra of the timing residuals for
eight pulsars are given in Figs.~\ref{f:plc} and \ref{f:plc2}.  Four of
the pulsars are those discussed in this paper and the remaining four are
discussed in Camilo~et~al. (in prep.). The spectra as a function
of frequency were converted to a function of period, while the timing signal
was obtained using $\Delta T_{a} = 2 \times 2 (P_{a}/N_{0})^{0.5}$, where
$P_{a}$ is the spectral power and $N_{0}$ is the number of independent data
points \cite{sca82}. Equation \ref{e:pmass} was then used to convert to
units of $m_{c} \sin i$. To account for the covariance with the period
derivative of the pulsar, the limits were multiplied by $P_{b}^{3}$ for
periods greater than one year.

To test the sensitivity of the method, the eccentricity of PSR~J1643$-$1224
was adjusted by 0.24 per cent which is precisely three times the uncertainty
in the measured value shown in Table \ref{t:par}.  The resulting
Lomb-Scargle spectrum in the lower panel of Fig.~\ref{f:plc} shows a highly
significant detection.  The probability that this signal is due to random
noise is $ 1.5\times 10^{-5}$. As another test of this method, the Jodrell
Bank timing residuals for PSR~B1257+12 were analysed in the same manner.
After fitting for a slow-down model of the pulsar, the spectrum of the
residuals (bottom panel of Fig.~\ref{f:plc2}) shows highly significant
signals at the periods of the planets. The probability
that these spikes are due to random noise is $ 2.8\times 10^{-8}$.  It is
important to note that the side lobes resulting from the uneven sampling of
the data are very strong at all periods. For example, in the B1257+12
spectrum, the rms is totally dominated by the side lobes, while the true
noise level for this pulsar is similar to that of the other pulsars shown in
Figs.~\ref{f:plc} and \ref{f:plc2}. This slightly increased rms may also be
seen by comparing the spectrum for PSR~J1643$-$1224 with the fake spectrum
below it in Fig.~\ref{f:plc}.

There are no spectral features that give any hint that a detectable periodic
signal is present in the residuals of any of the eight MSPs considered.
Probabilities of 0.1 and 0.999, that the signals are due to random noise are
shown by the smooth curves in the figures, indicating that to a very high
level of significance, the timing residuals are purely random.  For the
eight MSPs discussed here, companions with masses and orbits similar to
Venus, Earth and the two largest planets around PSR~B1257+12 are ruled out.
For those MSPs with more precise timing results, such as PSR~J0437$-$4715,
the existence of companions with masses and orbits similar to Mercury and
Mars are also ruled out.

\section*{Acknowledgments}
The authors are grateful to the referee Alex Wolszczan for helpful comments
and suggestions.


\clearpage

\begin{table*}
\caption{\label{t:par} Parameters of millisecond pulsars J0437$-$4715,
J0613$-$0200, J1045$-$4509, and J1643$-$1224.  Figures in parentheses
represent $1 \sigma$ uncertainties in the last significant digits
quoted.  Some values of $T_0$ and $\omega$ are given to greater
precision than the available accuracy to help observers. }

\begin{center}
\begin{tabular}{lllll}
\hline
\hline
\rule{0mm}{6mm}\hspace*{3.3cm} & PSR~J0437$-$4715  & PSR~J0613$-$0200 & 
PSR~J1045$-$4509 & PSR~J1643$-$1224
\\[2mm]
\hline
$\alpha$ (J2000)\dotfill & 
$04^{\rm h}37^{\rm m}15\fs73501(2)$ &
$06^{\rm h}13^{\rm m}43\fs97317(14)$ & 
$10^{\rm h}45^{\rm m}50\fs1940(3)$ & 
$16^{\rm h}43^{\rm m}38\fs1548(2)$    
\\
$\delta$ (J2000)\dotfill & 
$-47^{\circ}15'08\farcs1527(2)$ &
$-02^{\circ}00'47\farcs072(6)$ & 
$-45^{\circ}09'54\farcs204(2)$ &
$-12^{\circ}24'58\farcs740(14)$                     
\\
$\mu_{\alpha}$ (mas\,yr$^{-1})$\dotfill &
121.3(2) &
1.5(30)  &
$-7(3)$  &
$7(3)$   
\\
$\mu_{\delta}$ (mas\,yr$^{-1})$\dotfill &
$-70.4(3)$ &
$-4(6)$  &
8(2)    &
$7(17)$ 
\\
$\pi$ (mas)\dotfill                     
&1(5)  &... & ...   & ...     
\\
$P$ (ms)\dotfill   & 
5.75745182325672(13) & 
3.0618440360787(6) & 
7.4742241050635(10)  &
4.6216414453446(9) 
\\
$\dot P$\dotfill          & 
$5.7296(8)  \times10^{-20}$ & 
$9.57(4)\times10^{-21}$ & 
$1.740(8)\times10^{-20}$ & 
$1.840(6) \times10^{-20}$                           
\\
$|\ddot P| ({\mbox{s}}^{-1})$\dotfill &
$<1.4 \times 10^{-30}$ &
$<5 \times 10^{-30}$  &
$<14 \times 10^{-30}$   &
$<8 \times 10^{-30}$  
\\
Epoch (MJD)\dotfill & 
49615.0 & 
49512.0 & 
49597.0 & 
49524.0     
\\
DM (cm$^{-3}$\,pc)\dotfill & 
2.649(4)  & 
38.7911(5)  & 
58.1833(18) & 
62.4133(4)  
\\
$\dot {\rm DM}$ (cm$^{-3}$\,pc\,yr$^{-1})$ \dotfill & 
...  & 
0.0003(4)   & 
$-0.0049(13)$ & 
0.0010(5)
\\
$P_b$ (d)\dotfill & 
5.741042355(3) & 
1.198512557(3)  & 
4.083529188(17)  & 
147.017395(2) 
\\
$x$ (s)\dotfill & 
3.3666820(4)     & 
1.091445(5)    & 
3.015128(4)     & 
25.072613(4)   
\\
$\dot{x}$ \dotfill &
$ 7.4(14) \times 10^{-14}$ &
$< 20 \times 10^{-14}$ &
$  9(14) \times 10^{-14}$ &
$< 16 \times 10^{-14}$ 
\\
$e$\dotfill  & 
$1.86(3) \times10^{-5}$ & 
$7(10)   \times10^{-6}$ & 
$2.4(3)  \times10^{-5}$ & 
$5.057(4)\times10^{-4}$  
\\
$\omega$\dotfill &
$2\fdg1811\pm0.72$         &
$43\fdg0440\pm36.5$ &
$226\fdg078715\pm10$ &
$321\fdg8578\pm0.04$   
\\
$T_0$ (MJD)\dotfill &
49615.852918$\pm$0.01 &
49512.409269$\pm$0.2 &
49598.199375$\pm$0.1  &
49577.972054$\pm$0.015  
\\
$\dot{P}_{b}$ \dotfill&
$< 2.2 \times 10^{-11}$ &... & ...   & ...     
\\
\hline
$\mu$ (mas\,yr$^{-1}$)\dotfill & 140.2(3) & 4(6) & $11(3)$ & 10(13) \\
$d$ (kpc)\dotfill              & 0.14    & 2.2    & 3.2      & $>$4.9    \\
$v$ (km\,s$^{-1}$)\dotfill     & 93(30) & 40(60) & $168(45)$ & ...   \\
\hline
RMS residual ($\mu$s)\dotfill & 1.1  & 15  & 10 & 14    \\
Frequencies in fit (MHz)\dotfill  & 
1520, 1940 \& 2320 &
400, 600 \& 1400 &
1400 & 
400, 600 \& 1400 
\\
Data span (MJD)\dotfill &
49113$-$50117 &
49030$-$50138 &
49078$-$50117 & 
49046$-$50139 
\\
\hline
\end{tabular}
\end{center}
 
\end{table*}

\end{document}